\documentclass[prb,preprint]{revtex4}
\usepackage{graphicx}
\usepackage{amsmath}
\usepackage{color}

\begin{document}

\title{Exact solutions of a particle in a box with a delta function potential: The factorization method}
\author{Pouria Pedram}
\email{pouria.pedram@gmail.com}
\affiliation{Plasma Physics Research Center, Science and Research Campus, Islamic Azad University, P.O. Box 14665--678, Tehran, Iran}
\author{M. Vahabi}
\affiliation{Department of Physics, Shahid Beheshti University, G. C., Evin, Tehran 19839, Iran}

\begin{abstract}
We use the factorization method to find the exact eigenvalues and eigenfunctions for
a particle in a box with the delta function potential
$V(x)=\lambda\delta(x-x_{0})$. We
show that the presence of the potential results in
the discontinuity of the corresponding ladder operators. The presence of the delta function potential allows us
to obtain the full spectrum  in the first step of the
factorization procedure even in the weak coupling limit
$\lambda\rightarrow 0$.
\end{abstract}

\maketitle

\section{Introduction}\label{sec1}
The time-independent form of Schr\"{o}dinger's equation in the presence of a potential $V$ can be written as
\cite{Griffith}
\begin{equation}\label{Schrodinger}
- \frac{\hbar^{2}}{2 m} \nabla^2 \psi + V \psi=E \psi,
\end{equation}
where  $\psi$ is the
wave function, $m$ is the mass of the particle, and $E$ is the
eigenenergy. Equation~\eqref{Schrodinger} can be solved exactly only for
a few potentials. A particle in a box and a particle in a delta function potential are
two well-known and instructive examples.\cite{Griffith} The former can be used
to describe quantum dots and quantum wells at low
temperatures,\cite{Shklovskii,Cardona} and the latter can be used as
a model for atoms and molecules.\cite{Lapidus}

The solution for a particle in a box with a delta function potential
has been investigated using a perturbative expansion in the
strength of the delta function potential $\lambda$.\cite{Bera}
Exact solutions have been obtained
for the weak ($\lambda\rightarrow 0$) and the strong
($1/\lambda\rightarrow 0$) coupling limits.\cite{Joglekara}

In this paper we discuss the solution for a particle in a box with a
delta function potential using the factorization method and show
that the presence of the delta function simplifies the factorization
procedure. In this way we find the full spectrum of the Hamiltonian
in the first step of the factorization method. We also show that
this result applies in the weak coupling limit $\lambda \rightarrow
0$. Note that if we put $\lambda = 0$ from the beginning, we need to
continue the factorization procedure to find each eigenvalue in each
step.

\section{Particle in a box with a delta function potential}\label{sec2}
\begin{figure}
\centering
\includegraphics[width=8cm]{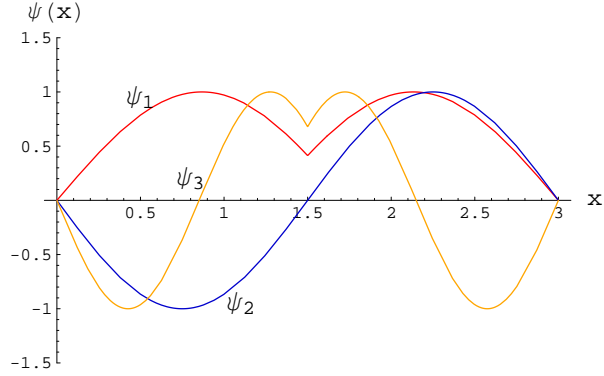}
\caption{First three stationary states $\psi_n(x)$ for
$2m\lambda/\hbar^2=8$, $a=3$, $A=1$, and $p=1/2$.} \label{fig1}
\end{figure}

Consider a particle in a one-dimensional box of size $a$ with
the delta function potential $V(x)=\lambda \, \delta(x-x_0)=\lambda
\, \delta(x-pa)$, where $0 < p < 1$. In this case
Eq.~\eqref{Schrodinger} takes the form
\begin{equation}\label{delta}
- \frac{\hbar^{2}}{2m} \frac{d^{2}\psi_n(x)}{d x^{2}} + \lambda
\delta(x-pa) \psi_n(x)= E_n \psi_n(x),
\end{equation}
where $\psi_n(x)$ and $E_n$ are the corresponding eigenfunctions and
eigenvalues, respectively. Because of the boundary conditions, $\psi_n(x)=0$ for $x\leq 0$ or $x\geq a$, the form of the eigenfunctions inside the box is
\begin{equation}\label{eigenfunction}
\psi_n(x) =
\begin{cases}
A \sin(k_nx) & (0 \leq x \leq pa) \\
B \sin[k_n(x-a)] & (pa \leq x \leq a),
\end{cases}
\end{equation}
where $k_n = \sqrt{\displaystyle 2mE_n}/\hbar$.
The continuity condition of the wave function at $x=pa$
gives $A/B =
\sin[(p-1)k_na]/\sin( pk_na)$. Because the delta
function is infinite at $x = pa$, the first
derivative of the wave function is not continuous and the relation
between the left and right derivatives of the wave function can be
obtained by integrating Eq.~(\ref{delta})
over the small interval $(x_0 - \epsilon, x_0 + \epsilon)$
\begin{subequations}
\begin{align}
\frac{d \psi_n(x)}{d x}|_{pa + \epsilon} - \frac{d \psi_n(x)}{d
x}|_{pa - \epsilon}
&=
\frac{2m}{\hbar^{2}}
\!\int_{pa - \epsilon}^{pa + \epsilon} V(x) \psi_n(x) \, dx \\
&=
\frac{2m\lambda}{\hbar^{2}}\psi_n(pa).\label{discontinuity1}
\end{align}
\end{subequations}
We substitute the eigenfunctions in Eq.~(\ref{eigenfunction}) in Eq.~(\ref{discontinuity1}) and obtain the
quantization condition\cite{Bera,Joglekara}
\begin{equation}\label{kequation}
k_n\sin(k_na)=\frac{2m\lambda}{\hbar^{2}}\sin(pk_na)\sin[(p-1)k_na].
\end{equation}
The solutions to Eq.~\eqref{kequation} give the energy spectrum of the
Hamiltonian,
$E_n=\hbar^2k_n^2/2m$. Figure~\ref{fig1} shows the first three stationary states for
$2m\lambda/\hbar^2=8$, $a=3$,
$A=1$, and $p=1/2$.

\section{The factorization method}\label{sec3}
To calculate the eigenvalues and eigenfunctions of a Hamiltonian
operator $H$, we can use a general operational procedure called the
factorization method. In this method the Hamiltonian is written as
the product of two ladder operators plus a constant,
$H=a^{\dag}a+E$. These operators are used to obtain the
eigenfunctions of the Hamiltonian. In contrast to the simple
harmonic oscillator, one ladder operator is usually not sufficient
to form all the Hamiltonian's eigenfunctions and  another ladder
operator is needed for each eigenfunction.

This method was first introduced by
Schr\"{o}dinger\cite{Schrodinger1,Schrodinger2,Schrodinger3} and
Dirac\cite{Dirac} and was further developed by Infeld and
Hull\cite{Infeld} and Green.\cite{Green} The spirit of the
factorization method is to write the second-order differential
operator $H$ as the product of two first-order differential
operators $a$ and $a^{\dag}$, plus a real constant $E$. The form of
these operators depends on the form of the potential $V(x)$ and the
factorization energy.

The procedure for finding the ladder operators and the eigenfunctions
is as follows.\cite{ohanian} We find operators $a_{1},
a_{2}, a_{3}, \ldots$ and real constants $E_{1}, E_{2}, E_{3}, \ldots$ from the recursive relations\cite{ohanian}
\begin{subequations}
\label{factorization1}
\begin{align}
a^{\dag}_{1}a_{1}+E_{1}&=H,\\
a^{\dag}_{2}a_{2}+E_{2}&=a_{1}a^{\dag}_{1}+E_{1},\\
a^{\dag}_{3}a_{3}+E_{3}&=a_{2}a^{\dag}_{2}+E_{2}, \ldots\,.
\end{align}
\end{subequations}
More generally
\begin{equation}\label{factorization2}
a^{\dag}_{n+1}a_{n+1}+E_{n+1}=a_{n}a^{\dag}_{n}+E_{n},
\qquad (n=1,2,\ldots).
\end{equation}
Also assume that there
exists a null eigenfunction (root function) $|\xi_{n}\rangle$ with
zero eigenvalue for each $a_{n}$, namely,
\begin{equation}\label{eigenvector1}
a_{n}|\xi_{n}\rangle=0.
\end{equation}
Hence, $E_{n}$ is the $n$th eigenvalue of the Hamiltonian with
the corresponding eigenfunction\cite{ohanian} (up to a
normalization coefficient)
\begin{equation}\label{eigenvector2}
|E_{n}\rangle=a^{\dag}_{1}a^{\dag}_{2}\ldots a^{\dag}_{n-1}|\xi_{n}\rangle.
\end{equation}

\begin{figure}
\centering
\includegraphics[width=8cm]{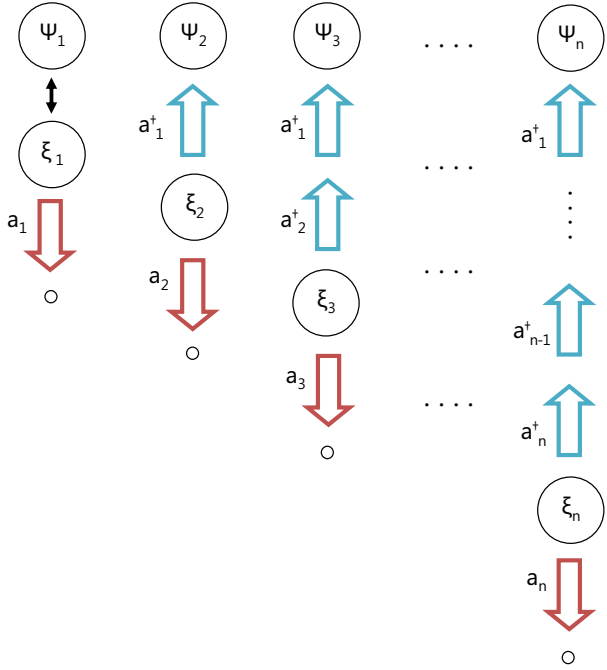}
\caption{A schematic diagram describing the relation between the
root functions $\xi_n$, the stationary states $\Psi_n$, and the
ladder operators $a_n$, $a_n^{\dagger}$.} \label{fig2}
\end{figure}

Figure \ref{fig2} shows a schematic diagram describing the relation
between root functions, the stationary states, and the ladder
operators. As can be seen from Eq.~(\ref{factorization1}),
$a^{\dag}_{1}a_{1}$ is equal to $H$ except for a constant. Thus,
$a_{1}$ must have a linear momentum term to be consistent with the
kinetic energy part of the Hamiltonian. According to Eq.~(\ref{factorization2}), each of the annihilation operators
$a_{n}$ should also have a a linear momentum term. Thus, $a_{n}$
can be written as
\begin{equation}\label{fj1}
a_{n}=\frac{1}{\sqrt{2m}}[P+i f_{n}(x)],
\end{equation}
where $P$ is the momentum operator and $f_{n}(x)$ is a real function
of $x$. Although these operators are not hermitian
($a^{\dag}_{n}=\dfrac{1}{\sqrt{2m}}(P-i
f_{n}(x))\ne a_{n}$), their product is hermitian
\begin{equation}\label{fj2}
a^{\dag}_{n}a_{n}=\frac{1}{2m}P^{2}+\frac{1}{2m}f_{n}^{2}+\frac{\hbar}{2m}
\frac{df_{n}}{dx}.
\end{equation}

We are now ready to find the ladder operators and eigenenergies for
our problem. First, we consider Eq.~(\ref{factorization1}) for
$n=1$
\begin{equation}
\label{above}
a^{\dag}_{1}a_{1}+E_{1}=H.
\end{equation}
From the form of the Hamiltonian (\ref{delta}) and
Eq.~(\ref{fj2}) we can rewrite Eq.~\eqref{above} as
\begin{equation}
\frac{P^{2}}{2m}+\frac{1}{2m}f_{1}^{2}+\frac{\hbar}{2m}
\frac{df_{1}}{dx}+E_{1}=\frac{P^{2}}{2m}+\lambda \delta (x-pa),
\end{equation}
or equivalently
\begin{equation}\label{delta2}
\frac{1}{2m}f_{1}^{2}+\frac{\hbar}{2m}
\frac{df_{1}}{dx}+E_{1}=\lambda \delta (x-pa).
\end{equation}
Note that in contrast to Eq.~(\ref{delta}), Eq.~\eqref{delta2} is a nonlinear first-order
differential equation.

To solve Eq.~(\ref{delta2}), we consider
the left-hand and right-hand sides of the delta function potential
separately. For these regions, Eq.~\eqref{delta2} reduces to
\begin{equation}
\frac{1}{2m}f_{1}^{2}+\frac{\hbar}{2m}
\frac{df_{1}}{dx}+E_{1}=0 \qquad (x\neq pa),
\end{equation}
which has the solution
\begin{equation}\label{f1withoutdelta}
f_{1}(x)=\sqrt{2mE_{1}}\cot[\frac{\sqrt{2mE_{1}}}{\hbar}(x-b)],
\end{equation}
where $b$ is a constant of integration. We require that $f_{1}(x)$
be finite in the range $0<x<a$, where it is the solution for our
problem. Because the singularities of the cotangent function are
$\pi$ radian apart, we choose the points $x=0$ and $x=a$ as the
singularity points of $f_{1}(x)$. At these points the
potential and hence the Hamiltonian are infinite and $f_{1}(x)$ can
be infinite at the boundaries. Hence, we have
\begin{equation}\label{f1withdelta}
f_{1}(x)=
\begin{cases}
\sqrt{2mE_{1}}\cot[\frac{\sqrt{2mE_{1}}}{\hbar}x], & (x<pa) \\
\sqrt{2mE_{1}}\cot[\frac{\sqrt{2mE_{1}}}{\hbar}(x-b)], & (x>pa),
\end{cases}
\end{equation}
where
$(\sqrt{2mE_{1}}/\hbar)(a-b)=\pi$.
To fix the value of $E_{1}$, we need to use the discontinuity
relation of the ladder operators, which can be obtained by integrating
Eq.~(\ref{delta2}) over a small interval ($pa-\epsilon,pa+\epsilon$)
\begin{equation}\label{discontinuity}
f_{1}(pa+\epsilon)-f_{1}(pa-\epsilon)=\frac{2m\lambda}{\hbar}.
\end{equation}
Equation~\eqref{discontinuity} shows that the presence of the delta function
results in the discontinuity of $f_{1}(x)$ at $x=pa$. We
use Eq.~(\ref{f1withdelta}) and $b=a-\pi
\hbar\sqrt{2mE_{1}}$, to rewrite Eq.~(\ref{discontinuity})
as
\begin{equation}
\label{above3}
\sqrt{2mE_{1}}\left\{\cot\Big[\frac{\sqrt{2mE_{1}}}{\hbar}(pa+\frac{\pi
\hbar}{\sqrt{2mE_{1}}}-a) \Big]-\cot \Big[\frac{\sqrt{2mE_{1}}}{\hbar}(pa) \Big]\right\}=\frac{2m\lambda}{\hbar}.
\end{equation}
Because $\cot (\pi + \alpha)=\cot \alpha$,
Eq.~\eqref{above3} can be written in the form
\begin{align}\label{disconfordelta}
&k_1\left\{\cot[(p-1)k_1a]-\cot(pk_1a)\right\} =\frac{2m\lambda}{\hbar^{2}}, \\
\noalign{\noindent or}
\label{spectrum}
&k_1\sin(k_1a)=
\frac{2m\lambda}{\hbar^{2}}\sin(pk_1a)\sin[(p-1)k_1a],
\end{align}
where $k_1=\sqrt{2mE_1}/\hbar$. Equation (\ref{disconfordelta}) is
similar to Eq.~(\ref{kequation}). Note that, although we calculated
the ground state energy of the Hamiltonian in the first step of the
factorization method, we found the full spectrum. This conclusion
applies to the weak coupling limit $\lambda\rightarrow0$. In this
limit $\sin(k_1a)=0$ from Eq.~(\ref{spectrum}), which gives the full
spectrum of a particle in a box with $k_1= n\pi/a$. This result has
an interesting consequence. If we use the factorization method to
calculate the energy levels of a particle in a box with $\lambda=0$,
we cannot find all of them in the first step of the procedure. In
fact, we should use the recursive relations to find each eigenvalue
in each step.\cite{ohanian} Thus, contrary to the Schr\"{o}dinger
equation, the presence of the delta function in the factorization
method simplifies the problem and repeated application of the
recursion relations is not necessary. We need only to replace the
index $1$ with $n$ in Eqs.~(\ref{f1withdelta}) and (\ref{spectrum}).

To obtain the root functions, we rewrite
Eq.~(\ref{eigenvector1}) as
\begin{subequations}
\label{above4}
\begin{align}
\left(\frac{\hbar}{i}\frac{d}{dx}+i \hbar k_n
\cot(k_nx)\right)\xi_{n}(x)&=0, \qquad (x<pa),\\
\left(\frac{\hbar}{i}\frac{d}{dx}+i \hbar k_n
\cot[k_n(x-b)]\right)\xi_{n}(x)&=0, \qquad (x>pa),
\end{align}
\end{subequations}
where $\xi_{n}(x)\equiv\langle x|\xi_{n}\rangle$. It can be easily
checked that the following is a solution to Eq.~\eqref{above4}
\begin{equation}
\xi_{n}(x)=
\begin{cases}
A\sin(k_nx), & (x<pa), \\
-B\sin[k_n(x-b)], & (x>pa).
\end{cases}
\end{equation}
Because in the first step of the factorization procedure the root
function $\xi_{1}(x)$ is equal to the first eigenfunction
$\psi_1(x)$ (see Eq.~(\ref{eigenvector2}) and Fig.~\ref{fig2}) and we
have replaced the index $1$ by $n$, all the root functions
are equal to the eigenfunctions of the Hamiltonian, that is,
$\xi_{n}(x)=\psi_n(x)$. This result is similar to
Eq.~(\ref{eigenfunction}) for $x<pa$. Moreover, because
$k_nb=k_na-\pi$, we have
$\sin[k_n(x-b)]=\sin[k_n(x-a)+\pi]=-\sin[k_n(x-a)]$. So, these
eigenfunctions are equivalent to the solutions in
Eq.~(\ref{eigenfunction}).

\section{Suggested problems}\label{sec6}
\textit{Problem 1.} Use the recursive relations
(\ref{factorization1}) to show that $|E_n\rangle$ in Eq.~(\ref{eigenvector2}) is the eigenfunction of the Hamiltonian with
$E_n$ as its eigenvalue.

\textit{Problem 2.} Show that the eigenvalues
$E_1,E_2,E_3,\ldots$ form a monotonic increasing sequence, that is,
$E_1\leq E_2\leq E_3\leq \ldots\,$.

\textit{Problem 3.} Apply the factorization method to a
particle in a box with no delta function potential. In the first
step show that $k_1=\pi/a$. Continue the factorization procedure to
find all the eigenenergies and eigenfunctions. Hint: consider
$a_{n}=\dfrac{1}{\sqrt{2m}}\left[P+ic_{n}\cot(
d_{n}x)\right]$, where $c_{n}$ and $d_{n}$ are real constants.


\begin{thebibliography}{99}
\bibitem{Griffith}D. J. Griffiths, \textsl{Introduction to Quantum Mechanics} (Prentice Hall, Upper Saddle River, NJ, 2004).

\bibitem{Shklovskii}B. I. Shklovskii and A. L. Efros, \textsl{Electronic Properties of Doped Semiconductors} (Springer, New York, 1984), Chap. 1.

\bibitem{Cardona}P. Y. Yu and M. Cardona, \textsl{Fundamentals of Semiconductors} (Springer, New York, 2005), Chap. 9.

\bibitem{Lapidus}I. R. Lapidus, ``Relativistic one-dimensional hydrogen atom,'' Am. J. Phys. \textbf{51}, 1036--1038 (1983).

\bibitem{Bera}N. Bera, K. Bhattacharya, and J. K. Bhattacharjee, ``Perturbative and non-perturbative studies with the delta function potential,'' Am. J. Phys. \textbf{76}, 250--257 (2008).

\bibitem{Joglekara}Y. N. Joglekara, ``Particle in a box with a $\delta$-function potential: Strong and weak coupling limits,'' Am. J. Phys. \textbf{77}, 734--736 (2009).

\bibitem{Schrodinger1}E. Schr\"odinger, ``A method of determining quantum-mechanical eigenvalues and eigenfunctions,'' Proc. Roy. Irish Acad. Sect. A \textbf{46}, 9--16 (1941).

\bibitem{Schrodinger2}E. Schr\"odinger, ``Further studies on solving eigenvalue problems by factorization,'' Proc. Roy. Irish Acad. Sect. A \textbf{46}, 183--206 (1941).

\bibitem{Schrodinger3}E. Schr\"odinger, ``The factorization of the hypergeometric equation,'' Proc. Roy. Irish Acad. Sect. A \textbf{47}, 53--54 (1941) or arXiv:physics/9910003.

\bibitem{Dirac}P. M. Dirac, \textsl{The Principles of Quantum Mechanics} (Oxford University Press, Oxford, 1958).

\bibitem{Infeld}I. Infeld and T. E. Hull, ``The factorization method,'' Rev. Mod. Phys. \textbf{23}, 21--68 (1951).

\bibitem{Green}H. S. Green, \textsl{Matrix Methods in Quantum Mechanics} (Barnes \& Noble, New York, 1968).

\bibitem{ohanian}H. C. Ohanian, \textsl{Principles of Quantum Mechanics} (Prentice Hall, Englewood Cliffs, NJ, 1990).
\end{thebibliography}
\end{document}